\documentclass{PoS}

\pdfoutput=1

\usepackage{amsmath}
\usepackage{multirow}

\newcommand{\mr}[2]{\multirow{#1}{*}{#2}}

\newcommand{\lst}[2]{#1^3\!\!\times\!#2}

\newcommand{\pbp}{\langle\bar\psi\psi\rangle}

\title{Lattice QCD with 8 and 12 degenerate quark flavors}

\ShortTitle{Lattice QCD with 8 and 12 degenerate quark flavors}

\author{\speaker{Xiao-Yong Jin}\\
        Department of Physics, Columbia University, New York, NY 10027, USA\\
        E-mail: \email{xj2106@columbia.edu}}

\author{Robert D. Mawhinney\\
        Department of Physics, Columbia University, New York, NY 10027, USA\\
        E-mail: \email{rdm@physics.columbia.edu}}

\abstract{

  We compare extensive simulations of QCD with 8 and 12 flavors of
  degenerate quarks, using the DBW2 gauge action, naive staggered
  fermions, and the rational hybrid Monte Carlo algorithm.  A variety
  of values of the coupling constant, quark mass, and lattice size
  have been used.  Our data suggests that, as the bare coupling is
  decreased, a rapid cross-over, which dramatically changes the
  lattice scale, exists with both 8 and 12 flavors.  The scale
  change across this cross-over is much larger with 12 flavors than
  it is with 8 flavors.  All of the observables we have measured,
  in both the zero and finite temperature systems, are consistent
  with a chiral symmetry breaking phase for the zero temperature
  theory on the weak coupling side of the rapid cross-over.

}

\FullConference{The XXVII International Symposium on Lattice Field Theory\\
  July 26 - 31, 2009\\
  Peking University, Beijing, China}

\begin{document}

%


\section{Zero temperature simulations}
\label{sec:zero-temp-simul}

In this paper, we report results from numerical simulations of QCD
with 12 flavors, which extend our previous investigation of 8-flavor
QCD \cite{Jin:2008rc}.  We seek evidence for the phase of zero
temperature QCD with these values of $N_f$, which is relevant
to locating a possible infrared fixed point and/or walking behavior.
We approach the problem directly by measuring standard hadronic
observables such as meson propagators and masses, the pion decay
constant, the static heavy-quark potential, and the string tension in
zero temperature lattice simulations.

As in our earlier, 8-flavor study \cite{Jin:2008rc}, we have used the
DBW2 gauge action and naive staggered fermion action for our study of
12-flavor QCD.  The DBW2 action reduces flavor symmetry breaking
\cite{Jin:2008rc} while still being fast in simulations.  The RHMC
algorithm is used here, even though we do not require fractional
powers of the fermion determinant, since this version of our code is
the most efficient.  Measurements are separated by 10 trajectories,
each of which has a length of 1/2 MD time units.  Additionally,
we have done an 8-flavor simulation with larger volume at
our smallest mass and weakest coupling to check finite volume effects
and have seen no changes.

\begin{table}
  \centering
  \begin{tabular}{l|l|cllll}
$\beta$       & $m_f$        & Size           & $\pbp$       & $m_\pi$     & $m_\rho$   & $\sigma$    \\
\hline
\hline
\mr{3}{0.45}  & 0.01         & $16^3\times32$ & 0.32521(26)  & 0.25327(51) &            &             \\
\cline{2-7}
              & 0.02         & $16^3\times32$ & 0.34687(17)  & 0.35376(34) &            &             \\
\cline{2-7}
              & 0.03         & $16^3\times32$ & 0.36290(16)  & 0.42867(28) &            &             \\
\hline
\mr{3}{0.46}  & 0.01         & $16^3\times32$ & 0.2687(17)   & 0.26430(47) &            &             \\
\cline{2-7}
              & 0.02         & $16^3\times32$ & 0.31240(57)  & 0.36137(35) &            &             \\
\cline{2-7}
              & 0.03         & $16^3\times32$ & 0.33586(12)  & 0.43617(22) &            &             \\
\hline
\mr{6}{0.47}  & \mr{3}{0.01} & $16^3\times32$ & 0.09047(51)  & 0.3159(11)  & 0.662(19)  & 0.0471(31)  \\
              &              & $24^3\times32$ & 0.09108(36)  & 0.31465(61) & 0.671(83)  & 0.0432(40)  \\
              &              & $32^3\times32$ & 0.09157(15)  & 0.31497(34) & 0.678(17)  & 0.04309(92) \\
\cline{2-7}
              & \mr{2}{0.02} & $16^3\times32$ & 0.25025(47)  & 0.37743(42) & 1.192(17)  & 0.377(39)   \\
              &              & $24^3\times32$ & 0.24995(37)  & 0.37787(35) & 1.2289(58) & 0.393(26)   \\
\cline{2-7}
              & 0.03         & $16^3\times32$ & 0.29695(25)  & 0.44691(31) & 1.295(46)  & 0.407(70)   \\
\hline
\mr{4}{0.475} & \mr{2}{0.01} & $16^3\times32$ & 0.07819(73)  & 0.3182(14)  & 0.625(17)  & 0.0446(22)  \\
              &              & $24^3\times32$ & 0.07651(23)  & 0.31721(79) & 0.5839(46) & 0.0315(12)  \\
\cline{2-7}
              & 0.02         & $16^3\times32$ & 0.1830(11)   & 0.39947(51) & 1.064(10)  & 0.100(21)   \\
\cline{2-7}
              & 0.03         & $16^3\times32$ & 0.26986(50)  & 0.45500(29) & 1.328(15)  & 0.372(58)   \\
\hline
\mr{6}{0.48}  & \mr{3}{0.01} & $16^3\times32$ & 0.06560(48)  & 0.3190(17)  & 0.5369(87) & 0.0279(26)  \\
              &              & $24^3\times32$ & 0.06662(12)  & 0.31253(41) & 0.5247(44) & 0.0224(14)  \\
              &              & $32^3\times32$ & 0.066839(61) & 0.31456(53) & 0.5308(32) & 0.02265(53) \\
\cline{2-7}
              & \mr{2}{0.02} & $16^3\times32$ & 0.13642(53)  & 0.41539(55) & 0.894(15)  & 0.0875(28)  \\
              &              & $24^3\times32$ & 0.13750(22)  & 0.41586(52) & 0.8952(94) & 0.0776(39)  \\
\cline{2-7}
              & 0.03         & $16^3\times32$ & 0.23246(55)  & 0.46619(45) & 1.1700(51) & 0.269(19)   \\
\hline
\mr{4}{0.49}  & \mr{2}{0.01} & $16^3\times32$ & 0.05326(67)  & 0.3293(37)  & 0.5231(75) & 0.0255(16)  \\
              &              & $32^3\times32$ & 0.05420(12)  & 0.30573(69) & 0.4497(37) & 0.0130(14)  \\
\cline{2-7}
              & 0.02         & $16^3\times32$ & 0.10335(34)  & 0.4230(10)  & 0.7353(60) & 0.0547(24)  \\
\cline{2-7}
              & 0.03         & $16^3\times32$ & 0.15994(45)  & 0.49419(41) & 1.006(13)  & 0.1056(55)  \\
\hline
\mr{3}{0.50}  & 0.01         & $16^3\times32$ & 0.04420(11)  & 0.3579(29)  & 0.5336(95) & 0.0189(12)  \\
\cline{2-7}
              & 0.02         & $16^3\times32$ & 0.08816(19)  & 0.4166(15)  & 0.6352(97) & 0.0399(22)  \\
\cline{2-7}
              & 0.03         & $16^3\times32$ & 0.12876(17)  & 0.49931(69) & 0.836(14)  & 0.0708(23)  \\
  \end{tabular}
  \caption{Simulation parameters and some results for 12-flavor QCD.}
  \label{tab:sim-param-12f}
\end{table}

Table \ref{tab:sim-param-12f} is a list of our major simulations with
12 flavors at $N_\tau = 32$.  We have done ordered and disordered
starts on the smallest lattices for all parameters, to check for a
unique phase for each parameter choice, and both starts produce the
same thermalized values for the chiral condensate, indicating no
metastability.  Generally 600-800 time units are used for
thermalization and then measurements are made every 10th time unit for
a total of $\sim 150$ measurements.  By simulating with multiple
volumes, we see that finite volume effects in $m_\pi$ and $m_\rho$ are
under control up to $\beta=0.475$, but that larger volumes are needed
at $\beta=0.49$.  The results from $\beta=0.50$ are not used in the
analysis here, because the volumes are too small.  The string tension
shows larger finite volume effects, which may be due to our
measurement method.  In this paper, results from the largest volume
available are used in the analysis.

\begin{figure}
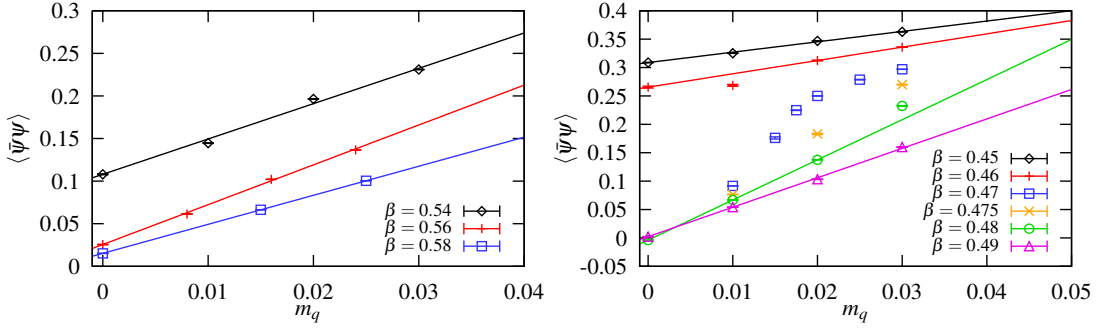

  \centering
  \includegraphics[width=0.47\textwidth]{fig/8f_pbp.mps}
  \includegraphics[width=0.47\textwidth]{fig/12f_pbp.mps}
  \caption{The chiral condensate for 8 flavors (left panel) with
    $\pbp(\beta=0.56,m_q=0)=0.02555(23)$.  The right panel is for
    12 flavors, where $\pbp(\beta=0.49,m_q=0)=0.00220(27)$}
  \label{fig:pbp}
\end{figure}

Fig.~\ref{fig:pbp} shows plots of the chiral condensate for both 8
and 12 flavors.  The 8-flavor graph on the left, from 3 equally
spaced $\beta$ values, shows a large change in $\pbp(m_q=0)$ with
$\beta$.  Linear extrapolations in $m_q$ give a non-zero value for
$\pbp(m_q=0)$, so the theory is in the chirally broken phase.  For
12 flavors (right graph), a transition from strong coupling to weak
coupling is seen, and in the transition region, non-linear quark
mass dependence is visible.  We see that for $\beta \lesssim 0.45$ the
system is clearly in a chirally broken phase, from the non-zero
value of $\pbp(m_q=0)$.  For $\beta \gtrsim 0.49$, we will argue that
the system is still in a chirally broken phase, but with a much
smaller scale (in lattice units) than the theory with $\beta \lesssim
0.45$.

We will refer to the region between $\beta \simeq 0.54$ and $\simeq
0.58$ (for 8 flavors) as $\Delta \beta_{8}$ and the region from $\beta
\simeq 0.46$ and $\simeq 0.49$ (for 12 flavors) as $\Delta
\beta_{12}$.  In this range of $\beta$ values, both systems are
showing rapid changes, and we want to determine the phase of the
theory on the weak coupling side of this rapid evolution region.

Throughout this report, we will be using simple analytic extrapolations
to $m_q = 0$.  For QCD-like theories in the chirally broken phase,
there will be chiral logarithms in such an extrapolation. {\em A
priori} one does not have any understanding of the quark mass range
over which chiral logs might be seen, since this requires knowing
the size of the chiral limit decay constant, $f$, and the chiral
condensate.  In addition, explicit factors of $N_f$ multiply chiral
logarithms, further influencing the quark mass range where they are
visible.  In 2+1 flavor QCD, where much work has been done on this
topic, simple linear extrapolations are rather accurate and we will
employ them here, but it is important to remember that we are making
this ansatz.

\begin{figure}
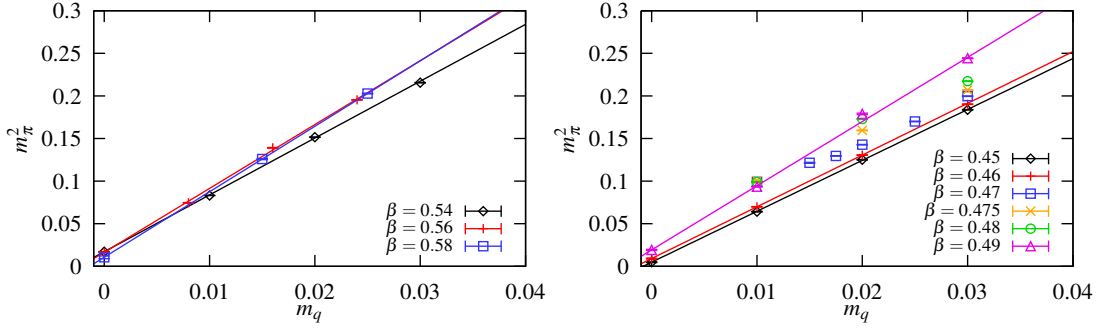

  \centering
  \includegraphics[width=0.47\textwidth]{fig/8f_pion_ext.mps}
  \includegraphics[width=0.47\textwidth]{fig/12f_pion_ext.mps}
  \caption{A plot of $m_\pi^2$ versus $m_q$ for 8 flavors (left panel)
    with $m_\pi^2(\beta=0.56,m_q=0)=0.01648(51)$.  The right panel is
    for 12 flavors with $m_\pi^2(\beta=0.45,m_q=0)=0.00482(39)$ and
    $m_\pi^2(\beta=0.49,m_q=0)=0.01926(66)$.}
  \label{fig:mpsquared}
\end{figure}

Fig.~\ref{fig:mpsquared} shows $m_\pi^2$ versus $m_q$.  For 8 flavors,
we can see clearly that $m_\pi^2 = 2B m_q$, the behavior expected for a
Goldstone boson.  $B$ changes little with $\beta$ and the intercept is
small.  (It is important to include chiral logarithms in an extrapolation
to test that $m_\pi$ = 0
when $m_q = 0$.) For 12 flavors, at strong coupling, the same
Goldstone behavior is seen.  In the $\Delta\beta_{12}$ region the
situation is not as clear.  For weaker coupling, $\beta = 0.49$, we
see noticeable finite volume effects in $m_\pi$ at the smallest quark
mass, $m_q = 0.01$.  (Finite volume effects are known to effect the
extrapolation of $m_\pi^2$ to $m_q = 0$ \cite{Sui:2001rf}.)  Since we
have a larger volume for $m_q = 0.01$, we use this result for $m_\pi$
in our fit and this gives a small intercept for $N_f = 12$ as shown in
the right panel of Fig.~\ref{fig:mpsquared}.  Thus we have a good
evidence for the pion, with 12 flavors, to be a Goldstone boson.  This
implies a non-zero dimensionful parameter, $B$, which also argues
against a conformal phase.  We note that for 12 flavors, $B$ also
has a mild dependence on $\beta$.

\begin{figure}
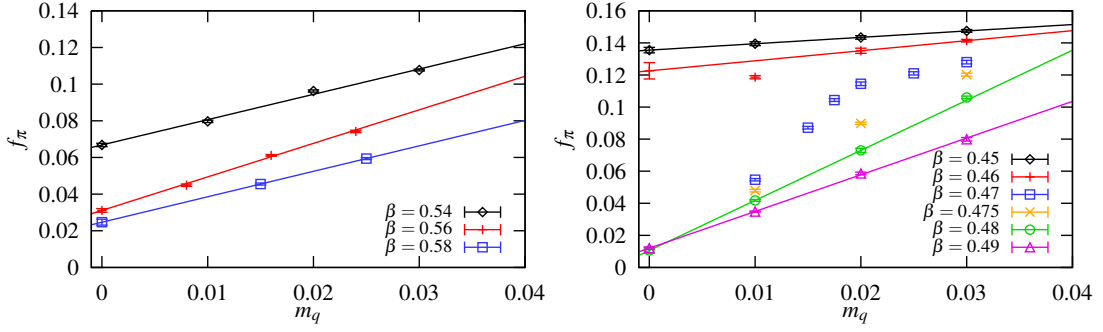

  \centering
  \includegraphics[width=0.47\textwidth]{fig/8f_fpi.mps}
  \includegraphics[width=0.47\textwidth]{fig/12f_fpi.mps}
  \caption{The pion decay constant for 8 flavors (left panel) and for
  12 flavors (right panel).}
  \label{fig:fpi}
\end{figure}

The pion decay constant, $f_\pi$, for 8 and 12 flavors is shown in
Fig.~\ref{fig:fpi}.  We can see that $f_\pi$ in the chiral limit
changes by about about $2\times$ for 8 flavors, across the
$\Delta\beta_8$ region, and by about $10 \times$ for 12 flavors.
Again we see evidence for a non-zero dimensionful parameter, $f_\pi$,
on the weak coupling side of the $\Delta\beta_{12}$ region.  We can
use the Gell-Mann--Oakes--Renner relation, $\pbp \propto m_\pi^2
f_\pi^2/ m_q$, to predict $\pbp$ for 12 flavors.  We have seen that
$m_\pi^2/m_q$ is almost independent of $\beta$, which means that $
\pbp \propto f_\pi^2$ on both sides of the $\Delta\beta_{12}$ region.
This gives a $100\times$ change in $\pbp$ across the $\Delta\beta_{12}$
region.  This is consistent with our direct measurements of $\pbp$
from Fig \ref{fig:pbp}, since the $100\times$ change in $\pbp$ makes
it likely to be too small to determine well in our current simulations.

\begin{figure}
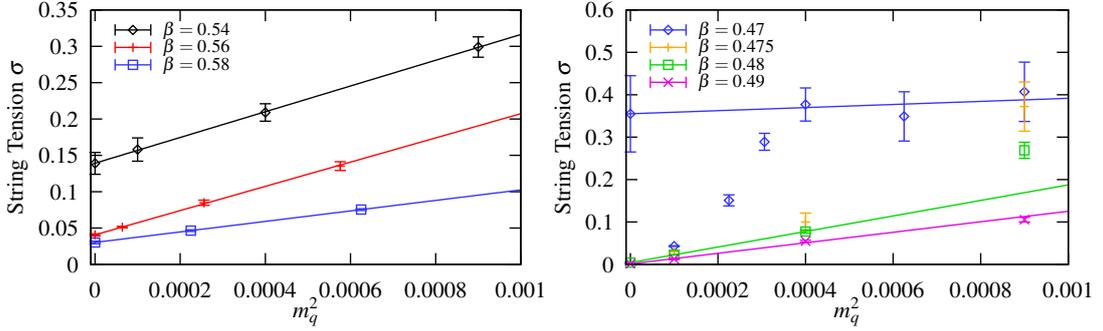

  \centering
  \includegraphics[width=0.47\textwidth]{fig/8f_sigma.mps}
  \includegraphics[width=0.47\textwidth]{fig/12f_sigma.mps}
  \caption{The string tension versus $m_q^2$ for 8 flavors (left panel)
    and 12 flavors (right panel), with
    $\sigma(\beta=0.48,m_q=0)=0.0043(15)$, and
    $\sigma(\beta=0.49,m_q=0)=0.0014(17)$.}
  \label{fig:sigma}
\end{figure}

We can also investigate the scale change in the string tension,
$\sigma$, as shown in Fig.~\ref{fig:sigma}.  Being a dimension two
quantity, $\sigma$ should scale as 
$f_\pi^2$.  It is easily seen for 8 flavors to change by roughly
$4\times$.  For 12 flavors, Fig.~\ref{fig:sigma} shows $\sigma
\approx 0.35$ in the chiral limit for $\beta=0.47$, found by linearly
fitting the heavier quarks with $m_q=0.02,0.025,0.03$.  (Lighter
quarks are in the rapid transition region and measuring $\sigma$
at smaller $\beta$ values is difficult because it is large.)  A
$10\times$ change in scale would give $\sigma \approx 0.0035$ on
the weak coupling side of the $\Delta\beta_{12}$ region, which is
consistent, within errors, with what we can measure.

\begin{figure}
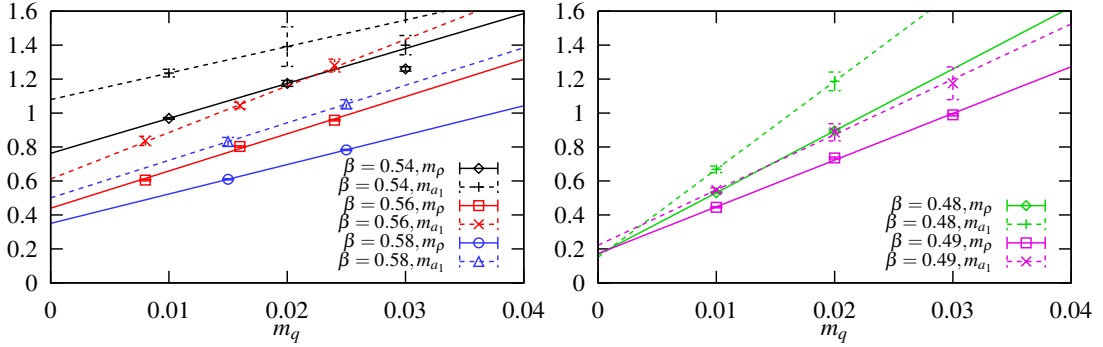

  \centering
  \includegraphics[width=0.47\textwidth]{fig/8f_rho_a1.mps}
  \includegraphics[width=0.47\textwidth]{fig/12f_rho_a1.mps}
  \caption{The spectrum of the parity partners, $\rho$ and $a_1$.  The
  left panel is for 8 flavors and the right for 12 flavors.  Dashed lines are
    linear extrapolation of $m_{a_1}$.  Solid lines are linear
    extrapolation of $m_\rho$.}
  \label{fig:rho-a1}
\end{figure}

It was observed \cite{Mawhinney:2000fw} that, for 4 flavors, the
spectrum of parity partners, like the $\rho$ and $a_1$, becomes
degenerate if the volume is too small, even if the theory is still in
the broken chiral symmetry phase.  The masses of the $\rho$ and $a_1$
are shown in Fig.~\ref{fig:rho-a1}, where, for 12 flavors, only values
at the two weakest coupling are shown, because the larger masses that
result from stronger coupling are hard to extract.  8-flavor QCD shows no
visible parity doubling, while for 12 flavors parity doubling is clearly
visible, suggesting finite volume effects are present in our simulations
of the chirally broken 12 flavor theory.

\section{Finite temperature simulations}
\label{sec:finite-temp-simul}

\begin{figure}
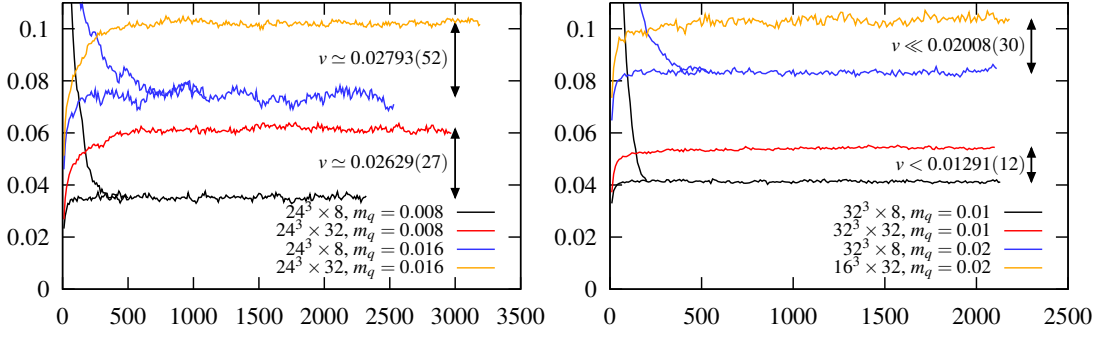

  \centering
  \includegraphics[width=0.47\textwidth]{fig/8f_pbp_nt8.mps}
  \includegraphics[width=0.47\textwidth]{fig/12f_pbp_nt8.mps}
  \caption{A comparison of the evolutions of $\pbp$ between $N_\tau=8$ and
    $N_\tau=32$.  The left panel is for 8 flavors and the right panel is for
    12.  Both ordered and disordered starts are shown for $N_\tau=8$, but
    only an ordered start is shown for $N_\tau=32$.}
  \label{fig:pbp-evo-finite}
\end{figure}

We have presented considerable evidence that 12 flavors is in a
chirally broken phase.  A check of these arguments is to run at higher
temperature and look for evidence of chiral symmetry restoration.
After the conference, we did finite temperature simulations at
$\beta=0.56$ with a lattice size of $\lst{24}{8}$ for 8 flavors, and
at $\beta=0.49$ with a lattice size of $\lst{32}{8}$ for 12
flavors. The evolutions of $\pbp$ are shown in
Fig.~\ref{fig:pbp-evo-finite}.  The amount the chiral condensate drops
at finite $T$ should be equal to the $T=0$ vacuum expectation value
which we have determined.  This is easily seen to be the case for 8
flavors, but for 12 flavors, we do not have an accurate measurement of
this $T=0$ quantity.  Also, with 12 flavors, the finite temperature
change in $\pbp$ varies with quark mass, consistent with the $T=0$
simulations not being entirely on the weak coupling side of the
$\Delta\beta_{12}$ region.

\begin{figure}
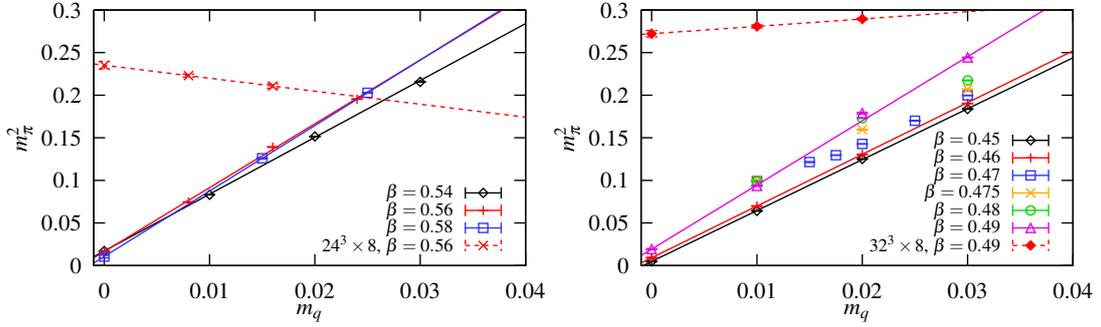

  \centering
  \includegraphics[width=0.47\textwidth]{fig/8f_mpi_sq_ext_nt8.mps}
  \includegraphics[width=0.47\textwidth]{fig/12f_mpi_sq_ext_nt8.mps}
  \caption{$m_\pi^2$ versus $m_q^2$.  Dashed lines are screening mass
    at $N_\tau=8$.  Left panel is 8 flavors.  Right panel is 12
    flavors.}
  \label{fig:pion-screening}
\end{figure}

A clearer signal comes from the screening mass of the pion, measured
along a spatial direction with $N_\tau = 8$.  This is shown in
Fig.~\ref{fig:pion-screening}.  At fixed $\beta$, we have changed
$N_\tau$ from 32 to 8, and seen a clear, dramatic change in the
behvior of $m_\pi$ with $m_q$.  Obviously, it is no longer a Goldstone
boson.  We have also measured the Polyakov loop along the
temporal direction for both $N_\tau = 8$ and 32.  It is clearly non
zero at $N_\tau=8$, and vanishes within errors for $N_\tau=32$.  We
conclude that there is a thermal transition taking place between $N_\tau=8$
and $N_\tau=32$ for both 8 flavors and 12 flavors.


\section{Conclusion}
\label{sec:conclusion}

We have presented considerable evidence that QCD with 8 and 12 flavors is
in a chiral symmetry breaking phase.  We see a pion which behaves as a
Goldstone boson for $T=0$ and which does not for finite $T$.  Multiple
dimensionful quantities are found, with $m_q = 0$, when $T=0$,
inconsistent with a conformal phase.

\begin{figure}
  \centering
  \includegraphics[width=0.47\textwidth]{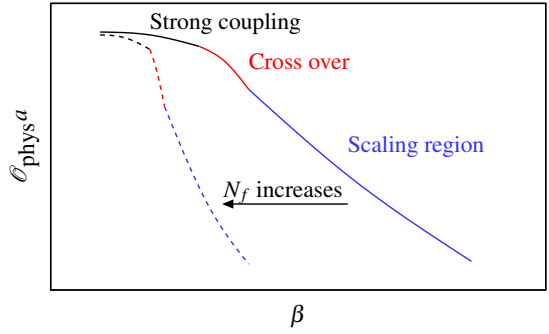}
  \caption{Generic dependence of an observable with $\beta$.}
  \label{fig:lattice}
\end{figure}

There is clearly a marked change in the system with 12 flavors, which
we discuss in the context of Fig.~\ref{fig:lattice}, which shows how
an observable in lattice units changes with $\beta$.  Previous $N_f =
8$ studies \cite{Brown:1992fz} argued that the cross-over could become
a phase transition, due to lattice artifacts.  With our DBW2 action,
this is no longer seen, rather we have a smooth, somewhat rapid,
change away from strong coupling.  This change becomes considerably
more rapid for $N_f = 12$, but the system still breaks chiral symmetry.
Currently, we do not know if our simulations at weaker coupling
are in the cross-over region, or at weak enough
coupling to begin to see the continuum, scaling region of the system.
Our rapid change for $N_f = 12$ could be an indicator of walking
behavior (as expected for this large number of flavors), or just a
steepened cross-over region still influenced by lattice artifacts.

We are thankful to all members of the RBC collaboration and especially
to Norman Christ for insightful discussions.  Our calculations were
done on the QCDOC at Columbia University and NY Blue at BNL.  This
research utilized resources at the New York Center for Computational
Sciences at Stony Brook University/Brookhaven National Laboratory
which is supported by the U.S. Department, of Energy under Contract
No. DE-FG02-92ER40699 and by the, State of New York.

\bibliographystyle{JHEP-2}
\bibliography{ref}

\end{document}